\documentclass[aps,superscriptaddress,twocolumn,showpacs,floatfix,notitlepage]{revtex4-2}

\usepackage{amsmath}
\usepackage[makeroom]{cancel}
\usepackage{amsfonts}
\usepackage{amssymb}
\usepackage{amssymb}
\usepackage{graphicx}
\usepackage{color}
\usepackage{xcolor}
\usepackage{xspace}
\usepackage[sort&compress]{natbib}
\usepackage{dsfont}
\usepackage{physics}
\usepackage[cal=boondox,scr=boondoxo]{mathalfa}

\newcommand{\excited}{{{$\raisebox{.04cm}{\bullet}\kern-.33cm=$}}}
\newcommand{\ground}{$\raisebox{-.04cm}{\bullet}\kern-.33cm=$}

\DeclareFontFamily{OT1}{pzc}{}
\DeclareFontShape{OT1}{pzc}{m}{it}%
{<-> s * [1.15] pzcmi7t}{}
\DeclareMathAlphabet{\mathpzc}{OT1}{pzc}{m}{it}

\newcommand{\ud}[1]{{#1^{\dagger}}}

\usepackage[colorlinks=true, linkcolor=blue, citecolor=magenta]{hyperref}

\allowdisplaybreaks

\begin{document} 
\title{Photon liquefaction in time}

\author{Eduardo {Zubizarreta Casalengua}}
\email{Eduardo.ZubizarretaCasalengua@wsi.tum.de}
\affiliation{Walter Schottky Institute, School of Computation, Information and Technology and MCQST, Technische Universit\"at M\"unchen, 85748 Garching, Germany}
\author{Elena {del Valle}}
\email{elena.delvalle.reboul@gmail.com}
\affiliation{Institute for Advanced Study, Technische Universit¨at M¨unchen, 85748 Garching, Germany}
\affiliation{Departamento de F\'isica Te\'orica de la Materia Condensada, Universidad Aut\'onoma de Madrid, 28049 Madrid, Spain}
\author{Fabrice P. Laussy}
\email{fabrice.laussy@gmail.com}
\affiliation{Faculty of Science and Engineering, University of Wolverhampton, Wulfruna St, Wolverhampton WV1 1LY, UK}

\date{\today}

\begin{abstract}
  We provide a mechanism to imprint local temporal correlations in
  photon streams which have the same character as spatial correlations
  in liquids. Usual single-photon emitters correspond, in this
  picture, to a (temporal) gas while uncorrelated light is the ideal
  gas. We argue that good single-photon sources are those that exhibit
  such temporal liquid features, i.e., with a plateau for their
  short-time correlations (as opposed to a linear dependence) and
  oscillations at later times, which is a direct manifestation of
  photon time-ordering. We obtain general, closed-form analytical
  expressions for the second-order coherence function of a broad
  family of ``liquid light'' which can be arbitrarily correlated,
  though never completely crystallized.
\end{abstract}
\maketitle	


A liquid is a condensed phase of matter whose definition at the
microscopic level has been the subject of much
debate~\cite{chandler17a}. It consists of a dense, disordered assembly
of molecules exhibiting short-range order in space. Simplest liquids
like monatomic argon are well modeled as jumbled closely packed
spheres, with an ordering governed by integer multiples of the
molecular diameter~\cite{finney13a}. The molecular arrangement can be
revealed experimentally by diffracting X-rays or neutrons on the
fluid~\cite{amannwinkel16a}. The absence of a Bragg peak indicates
that there is no long-range order, but oscillations in the radial
diffracted intensities reveal short-range correlations, whereby each
molecule is locally attached to a shell of its surrounding neighbors,
that remain free to move around and distort, but with more or less
probability to be at a given distance. A gas, in contrast, presents no
such correlations and has no short-range order. Two molecules still
cannot sit at the same position so there remains a depletion of
probabilities for close distances, but there are no oscillations. In
condensed-matter physics, this is described by the structure factor,
whose Fourier transform provides the so-called pair-correlation
function~$g(r)$ that yields the probability of finding a molecule at a
distance~$r$ from another molecule, relative to an
uncorrelated---i.e., ideal---gas~\cite{barker76a}. All these
correlations are in space.

Independently from these statistical considerations for correlations
of distances between molecules, quantum optics arrived at the modern
definition of quantum coherence of light through correlations of
photons in time~\cite{glauber63a}. This relies on the so-called
second-order coherence function~$g^{(2)}(\tau)$, with a notation
eerily reminiscent of the condensed-matter case, although, again,
there appears to have no trace of any connection from one field to the
other. The $g^{(2)}(\tau)$ function similarly quantifies the density
of two-photons separated by a time delay~$\tau$, as compared to an
uncorrelated (Poissonian or, as the optical terminology goes,
``coherent'') photon stream~\cite{glauber63c}.

There are obvious differences between the two cases: fluids are
typically three-dimensional and their correlations are in space, while
quantum optics treats with one-dimensional photon correlations in
time.  There are, however, more similarities than seems to have been
previously appreciated. In quantum optics, the most studied type of
quantum correlations is for single-photon sources~\cite{lounis05a},
with a suppression of two-photon coincidences, i.e., two photons are
never detected at exactly the same time. This is not trivial since
photons, being bosons, have the natural tendency of exhibiting the
opposite behavior of bunching. Some order must be imbued to the
photon stream to fight their urge of coming together.  The simplest
way to achieve this is to recourse to a two-level system~$\sigma$, put
in its excited state at a rate~$P_\sigma$. If the emitter has a
radiative decay rate~$\gamma_\sigma$, one finds for its second-order
coherence~\cite{grangier86a}:
\begin{equation}
  \label{eq:Sat22Jul164940CEST2023}
  g^{(2)}(\tau)=1-\exp\big(-(P_\sigma+\gamma_\sigma)\tau\big)\,.
\end{equation}
This has coherence time~$P_\sigma+\gamma_\sigma$ with a linear~$\tau$
short-time loss of coherence from perfect two-photon
suppression~$g^{(2)}(0)=0$ to uncorrelated emission
$\lim_{\tau\to\infty}g^{(2)}(\tau)=1$. Another paradigmatic type of
excitation is coherent excitation where a classical field (typically a
laser) drives resonantly the two-level system, bringing it in another,
much richer regime including quantum dressing of the transitions and
coherent scattering. In the high-driving regime~\cite{carmichael76a}:
\begin{equation}
  \label{eq:Mon24Jul170315CEST2023}
  g^{(2)}(\tau)=1-e^{-{3\over4}\gamma_\sigma\tau}\left[\cosh\left({\gamma_\mathrm{M}\tau\over4}\right)+{3\gamma_\sigma\over\gamma_\mathrm{M}}\sinh\left({\gamma_\mathrm{M}\tau\over4}\right)\right]
\end{equation}
where~$\gamma_\mathrm{M}\equiv\sqrt{\gamma_\sigma^2-(8\Omega_\sigma)^2}$
is the Mollow (also known as Rabi) splitting. In the low-driving,
so-called Heitler regime, when~$\Omega_\sigma\ll\gamma_\sigma$, the
two-photon correlations takes the simpler form:
\begin{equation}
  \label{eq:Sat22Jul171356CEST2023}
  g^{(2)}(\tau)=\big(1-\exp(-\gamma_\sigma\tau)\big)^2\,.
\end{equation}
In both cases (Eqs.~(\ref{eq:Mon24Jul170315CEST2023})
and~(\ref{eq:Sat22Jul171356CEST2023})), there is a qualitative change
of the short-time correlations---where photon suppression
occurs---from a linear~$\tau$ dependence in the incoherent case of
Eq.~(\ref{eq:Sat22Jul164940CEST2023}) to a quadratic~$\tau^2$ one for
the coherent cases. Such change from linear to power dependence on
time typically reflects a qualitative transformation of the response
of a system. For fluids, such dependencies in an energy spectrum are
for instance responsible for its diffusive or superfluid
character~\cite{landau41a}.

Although not still compelling at this point of our discussion, we
highlight that the two-photon correlation
function~(\ref{eq:Sat22Jul164940CEST2023}) exhibits oscillations
when~$\gamma_\mathrm{M}$ becomes imaginary, i.e.,
when~$\Omega_\sigma>\gamma_\sigma/8$, marking the onset of Mollow
Physics.  In this case, they are understood as Rabi oscillations of
the two-level system which gets dressed by the
laser~\cite{schaibley13a}. There is then a transition from a
monotonous~$g^{(2)}(\tau)=1-e^{-3\gamma_\sigma\tau/4}(1+\gamma_\sigma\tau/4)$
at threshold to one featuring all-times oscillations:
$g^{(2)}(\tau)=1-e^{-3\gamma_\sigma\tau/4}\big(\cos(\Omega_\mathrm{M}\tau/4)+{\gamma_\sigma\over\omega_\mathrm{M}}\sin(\Omega_\mathrm{M}\tau/4)\big)$
with~$\Omega_\mathrm{M}\equiv\sqrt{(8\Omega_\sigma)^2-\gamma_\sigma^2}$
a real parameter. The maximum
$g^{(2)}(\tau_\mathrm{M})=1+\exp(-3\pi\gamma_\sigma/\Omega_\mathrm{M})$
is obtained at~$\tau_\mathrm{M}=4\pi/\Omega_\mathrm{M}$ and thus is at
most~$2$, in the limit of~$\Omega_\sigma\to\infty$.  Such oscillations
are understood as Rabi oscillations of the populations since in this
case
$g^{(2)}(\tau)=n_0(\tau)/n_\mathrm{ss}$~\cite{arXiv_zubizarretacasalengua23b}
where~$n_0(\tau)$ is the dynamics of the system when starting from its
ground state and~$n_\mathrm{ss}$ is the steady state population, which
is at most~${1\over2}$ for the coherently driven system since
stimulated emission prevents population inversion.

While such a Rabi interpretation is completely
valid~\cite{diedrich87a,makhonin14a,konthasinghe19a}, in the
following, we shall argue that such familiar oscillations are a
particular case of a more general trend, namely, photon liquefaction
in time, by what we mean temporal ordering of the photons similar to
that in space when a gas becomes liquid.  The terminology of
``condensation'' is more common to describe gas-to-liquid transition,
but given the predominance of Bose condensation for bosons, we prefer
here to refer to that phenomenon with the alternative denomination of
``liquefaction'' which, we highlight again, further occurs in
time. This approach is motivated by the notion of a ``perfect single
photon source''~\cite{arXiv_khalid23a} understood as a source which
suppresses photon coincidences not only at exactly~$\tau=0$, but over
a temporal window large enough or robust enough so that a physical
detector will be resilient to the unavoidable time uncertainty
associated to the photodetection process. Such temporal limitations of
physical detectors were first highlighted for single-photon
observables by Eberly and W\'odkiewicz~\cite{eberly77a} and later
upgraded to multiphoton detection by del Valle \emph{et
  al.}~\cite{delvalle12a}.  For two-photon suppression, this results
in photons correlations of the type of
Eqs.~(\ref{eq:Mon24Jul170315CEST2023})
and~(\ref{eq:Sat22Jul171356CEST2023}) to be much more resilient to
time-frequency uncertainties and, correspondingly, to provide much
better antibunching and less ``accidental'' coincidences, due to the
flatter short-time correlation~$\tau^2$~\cite{lopezcarreno22a}. In
Ref.~\cite{arXiv_khalid23a}, it was shown that in the mathematical
idealization where the correlation is flattened so much as to actually
open a non-analytic time-gap, i.e., forbidding completely two photons
to be closer than a given time~$t_\mathrm{G}$, then oscillations ensue
in~$g^{(2)}(\tau)$ as a result of time-ordering, thus being a direct
counterpart, but in time, of the transition from a gas to a liquid. In
fact, for the case of a perfect, rigid time-gap, correlations are
precisely those, in space, for a system of hard rods, as was first
described by Prins~\cite{zernike27a} who also was the first to derive
the expression to compute diffracted intensities from molecular
arrangements.

This mathematically-perfect single photon
source~\cite{arXiv_khalid23a} was analyzed with no underlying physical
mechanism to realize it. Here, we provide a broad class of photon
temporal liquids, based on a simple mechanism whereby the excitation
undergoes a cascade of transitions between various states before
ultimately emitting a photon.  This is particularly relevant for
solid-state systems~\cite{aharonovich16a} where the two-level system
is implemented by an artificial atom embedded in an environment which
comes with various intermediate states, shell structures, metastable
states, etc., which could even be controlled or ultimately
engineered~\cite{faist94a}. We find the interesting result that even
\emph{incoherent} driving, insofar as it involves intermediate steps
in the cascade, can feature a two-photon correlation function that
corresponds to the liquid phase, with oscillations and a power-law
dependence for the short-times correlations.  The power is furthermore
directly related to the number of cascades, and produces the flat
plateau typical of hard-sphere repulsions in condensed matter, as well
as the characteristic bunching elbows that mark the onset of local
ordering~\cite{arXiv_khalid23a}.


\begin{figure}
  \includegraphics[width=.9\linewidth]{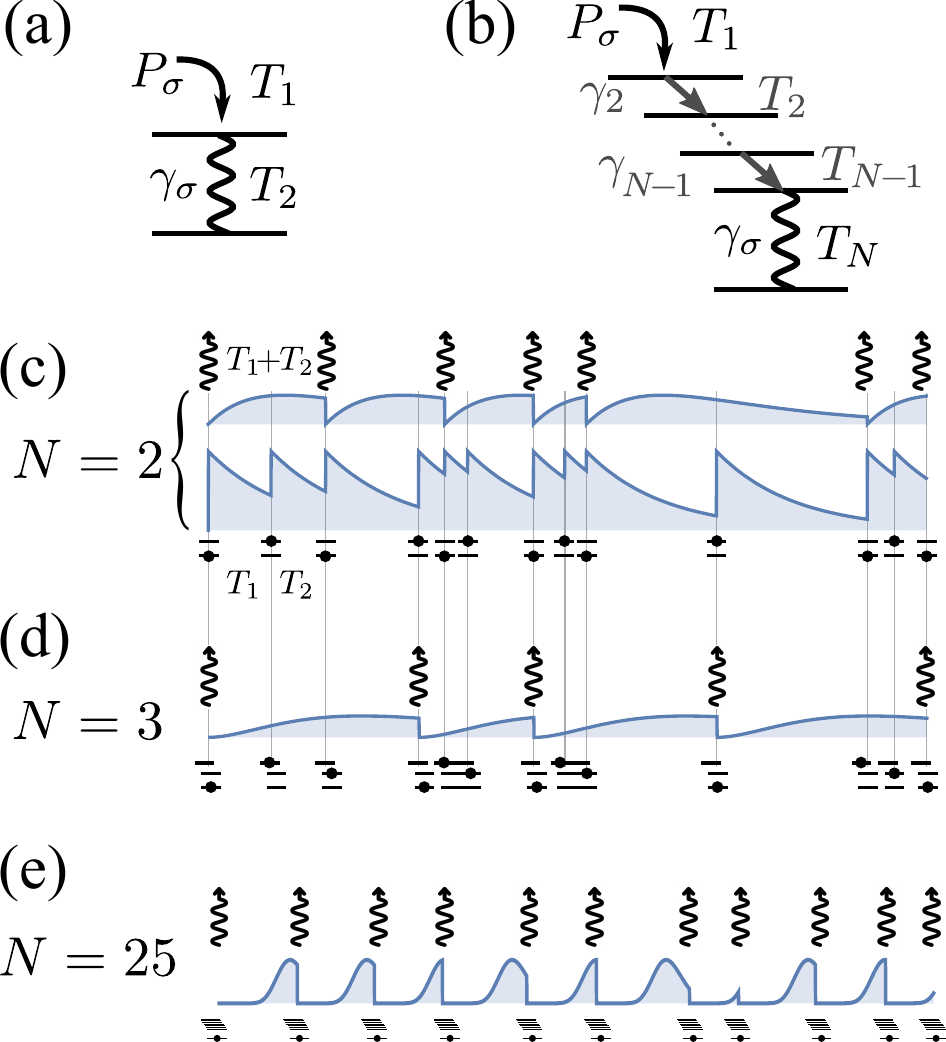}
  \caption{Single-photon emission from (a) a two-level system under
    incoherent pumping~$P_\sigma$ and decay~$\gamma_\sigma$ and (b)
    with a prior cascade of intermediate transitions. (c) Photon
    emissions from successive random processes of excitation and
    emission, described by exponential distributions or directly with
    the corresponding Erlang distributions for two events. (d) Case
    with~$N=3$ featuring one intermediate cascade and the
    corresponding Erlang distributions. (e) Case with~$N=25$ (with a
    rescaled time axis) with the corresponding Erlang distributions
    and only the final states (emission) of the cascade depicted. The
    photon stream is neatly ordered in time with short-range
    correlations of a liquid, although there are no interactions and
    all processes are incoherent.}
  \label{fig:Fri29Dec004818CET2023}
\end{figure}

\begin{figure}
  \includegraphics[width=\linewidth]{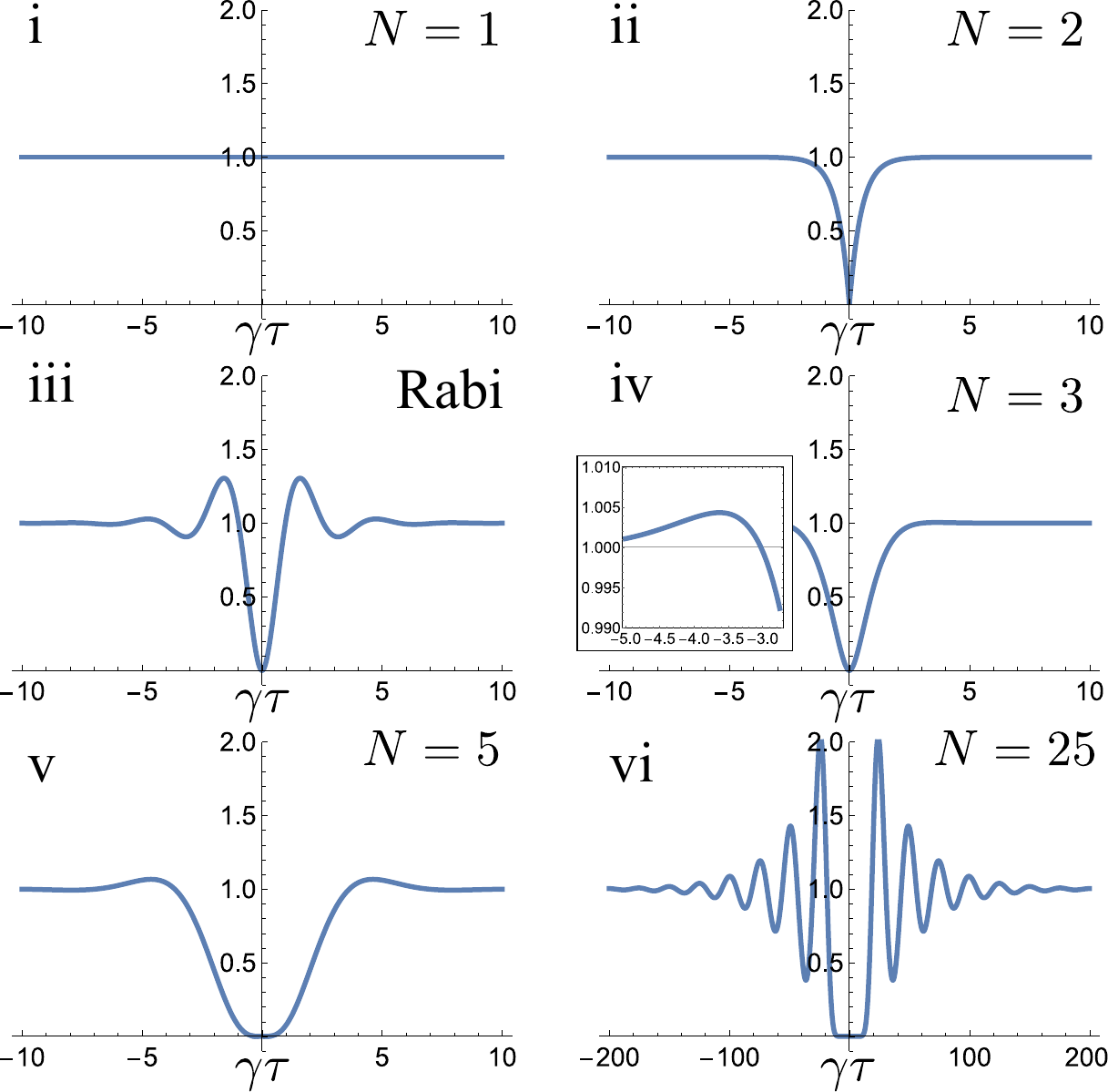}
  \caption{Two-photon correlation functions for i) coherent light
    (ideal gas) ii) an incoherently pumped 2LS (gas), iii) a strongly
    and coherently-driven 2LS (Rabi liquid), iv) an
    incoherently-driven one-cascade 2LS (onset of liquid with a zoom
    in inset showing the $g^{(2)}=1.0043$ bunching), v) liquefaction
    with~$5$ cascades and vi) liquid with~$25$ cascades. Note that
    the~$\tau$ axis is rescaled in the latter case as over the
    range~$\pm 10/\gamma$ of the other plots, $g^{(2)}$ remains below
    $2\times 10^{-3}$. We assumes all rates to be equal.}
    \label{fig:Thu27Jul141838CEST2023}
\end{figure}

As opposed to solving a quantum optical master equation---which yields
the same result, as we shall show for a particular case---we take a
more insightful statistical and condensed-matter-inspired approach
based on two-photon correlation functions and their underlying waiting
time distributions.  The incoherent two-photon
emission~(\ref{eq:Sat22Jul164940CEST2023}) can be understood as a
two-steps process with, first, an underlying, backbone Poisson
(uncorrelated) stream of events, corresponding to the incoherent
excitation at rate~$P_\sigma$ that brings the system in its excited
state. Each such event then draws another Poisson-distributed random
number with parameter~$\gamma_\sigma$, describing the spontaneous
emission (second step)~\cite{arXiv_khalid23a}. This is sketched in
Fig.~\ref{fig:Fri29Dec004818CET2023}(a) where the exponential
(Poisson) distributions alternate sampling of uncorrelated
excitation~$T_1$ and spontaneous emission~$T_2$ times.  Since we are
only interested in the emitted photons, we can equivalently consider
directly the distribution for~$T_1+T_2$, which is given by
$\tau\exp(-(P_\sigma+\gamma_\sigma)\tau)$. If the emission involves
$N$ steps, first with parameter~$P_\sigma$ and each subsequent one
with parameter~$\gamma_i$ ($2\le i\le N$), then one needs to similarly
replace the exponential decay by the distribution for the sum of $N$
independent exponential random variables, which is one of the
phase-type class of distributions, known as the Hypoexponential
distribution with $N$ parameters. When those are all equal, the
distribution is more popularly known as the Erlang distribution and we
shall focus on this case for conceptual simplicity.  The waiting time
distribution for~$N$ steps (one excitation plus $N-1$ cascades) can
then be simply obtained as:
\begin{equation}
  \label{eq:Sat22Jul194910CEST2023}
  w(\tau)={\gamma^N\tau^{N-1}e^{-\gamma\tau}\over(N-1)!}\,.
\end{equation}
Since two-photon correlation functions can be computed from the
waiting time distribution $w(\tau)$ by transiting to the Laplace
space~\cite{kim87a}
\begin{equation}
  \label{eq:Wed24May154248BST2023}
\tilde g^{(2)}(s)\equiv\int_0^\infty
g^{(2)}(\tau)e^{-s\tau}\,d\tau=\gamma{\tilde w(s)\over 1-\tilde w(s)}  \,,
\end{equation}
with~$\tilde w(s)=\gamma^N/[(s+\gamma)^N-\gamma^N]$ the Laplace
transform of Eq.~(\ref{eq:Sat22Jul194910CEST2023}), we then find, by
inverse Laplace transform:
\begin{equation}
  \label{eq:Tue25Jul142700CEST2023}
  g^{(2)}_N(\tau) = 1 + \sum_{p=1}^{N-1} z_N^p \exp\big(- \gamma (1- z_N^p)\tau\big)
\end{equation}
where~$z_N\equiv\exp\left(i{2\pi/ N}\right)$ 
%
are the $N$th roots of unity. This succinct general expression can be
easily made explicit for particular cases, e.g., with no cascade
($N=1$ for the excitation alone), we have an uncorrelated (or
coherent) photon stream, Fig.~\ref{fig:Thu27Jul141838CEST2023}(a),
while~$N=2$ describes the excitation plus spontaneous emission of
Fig.~\ref{fig:Fri29Dec004818CET2023}(c) and thus recovers
Eq.~(\ref{eq:Sat22Jul164940CEST2023}).  We get new results with two
cascades ($N=3$, Fig.~\ref{fig:Fri29Dec004818CET2023}(d)), for which
Eq.~(\ref{eq:Tue25Jul142700CEST2023}) simplifies to
\begin{equation}
  \label{eq:Sat22Jul195618CEST2023}
  g^{(2)}(\tau)=1-2\sin\left({\sqrt3\over 2}\gamma\tau+{\pi\over6}\right)e^{-{3\over2}\gamma\tau}
\end{equation}
and for three cascades ($N=4$):
\begin{equation}
  \label{eq:Mon24Jul143926CEST2023}
  g^{(2)}(\tau)=1-e^{-2\gamma\tau}-2e^{-\gamma\tau}\sin(\gamma\tau)\,
\end{equation}
with the possibility to derive similar closed-form expressions for
other~$N$.

These cascaded chains of incoherent relaxation produce, interestingly,
two-photon correlation functions that are more like in character those
of the coherently driven case~(\ref{eq:Mon24Jul170315CEST2023}) than
the incoherent case~(\ref{eq:Sat22Jul164940CEST2023}).  Even with one
cascade only, there is some onset of liquefaction with oscillations
which, although not compelling numerically, are clear from the
analytical expressions in Eq.~(\ref{eq:Sat22Jul195618CEST2023}).  The
maximum $g^{(2)}(\tau_\mathrm{M})=1+e^{-\sqrt 3\pi}\approx 1.0043$
at~$\gamma_\sigma\tau_\mathrm{M}=2\pi/\sqrt{3}$, however, is only
marginally different from unity. Higher~$g^{(2)}$ are obtained for a
higher number of cascades but at time delays which are zeros of
transcendental equations (for instance, as a solution of
$\exp(\tau)\sin(\tau+\pi/4)=c$ for~$N=2$) but these can be easily
obtained numerically.  Another manifestation of temporal liquefaction
is the hardening of the photons that increasingly repulse each other,
with short-time expansion from~$N$ cascades providing the plateaus:
\begin{equation}
  \label{eq:Tue25Jul153331CEST2023}
  g^{(2)}(\tau)\approx{N^2\over N!}(\gamma\tau)^N\,.
\end{equation}
This dependency of the short-time correlations mean that a physical
detector will be increasingly less affected, at a qualitative level,
by the fundamental time uncertainty attached to photodetection. These
qualities are directly inherited from the waiting time
distribution~(\ref{eq:Sat22Jul194910CEST2023}) and the shape of the
probability distribution for each photon emission.  The flattening of
these quantities at short times makes it possible to open an effective
temporal gap. This paves the way towards a genuine perfect
single-photon source, with no multiphoton emission for detectors with
a better temporal resolution than the time gap, just as a
superconductor is a genuine perfect conductor, with no loss
whatsoever. The peak following the plateau in the emission probability
also imprints a clear time-ordering of the photons, as a result of the
compound-time averaging out its more extreme fluctuations. Together,
these two features produce a regular stream of photons. This is
obvious in Fig.~\ref{fig:Fri29Dec004818CET2023}(e) where photons
appear to be equidistantly spaced in time with small only
fluctuations, as would be expected from an externally-controlled
(pulsed) single-photon source~\cite{arXiv_khalid23a}. Such an order is
local only, however, as fluctuations, however small, pile up and
eventually wash out correlations for photons distant enough. This thus
corresponds to a photon liquid, and never a crystal, as the stream is
intrinsically stationary for long-enough times. This is, however, of
little concern for the single-photon character.

While these properties are well-understood from such a point-process
statistical approach, which is furthermore rooted in statistical
arguments that have been thoroughly studied in condensed-matter
physics, these results would also be obtained from a quantum
treatment, i.e., with a master equation for a multilevel system. The
case~$N=3$, for instance, with level structure~$\ket{i}$
for~$0\le i\le 2$ and
Hamiltonian~$H\equiv\sum_{i=1,2}\omega_i\ketbra{i}{i}$ can be
described by the
Lindbladian~$\mathcal{L}\rho\equiv=-i[H,\rho]+\big\{{P_\sigma\over2}\mathcal{L}_{\ketbra{2}{0}}+{\gamma_1\over2}\mathcal{L}_{\ketbra{1}{2}}+{\gamma_2\over2}\mathcal{L}_{\ketbra{0}{1}}\big\}\rho$
where for any operator~$\Omega$,
$\mathcal{L}_\Omega\rho\equiv2\Omega\rho\ud{\Omega}-\ud{\Omega}\Omega\rho-\rho\ud{\Omega}\Omega$.
One can then compute steady-state two-time correlators
like~$G^{(2)}_{10}(\tau)\equiv\langle\ud{\sigma}_{10}(0)\ud{\sigma}_{10}(\tau)\sigma_{10}(\tau)\sigma_{10}(0)\rangle$
for $\sigma_{10}\equiv\ketbra{0}{1}$ from the quantum regression
theorem, i.e.,
$G_{10}^{(2)}(\tau)=\operatorname{Tr}\big(\ud{\sigma}_{10}\sigma_{10}e^{\mathcal{L}\tau}[\sigma_{10}\rho_\mathrm{ss}\ud{\sigma_{10}}]\big)$
on the steady-state density matrix which is diagonal:
\begin{equation}
  \label{eq:Fri29Dec183552CET2023}
  \rho_\mathrm{ss}={1\over\gamma_1\gamma_2+P_\sigma(\gamma_1+\gamma_2)}
  \begin{pmatrix}
    \gamma_1\gamma_2\\
    &P_\sigma\gamma_1\\
    &&P_\sigma\gamma_2
  \end{pmatrix}\,.
\end{equation}
The normalization of~$G^{(2)}_{10}$ yields~$g^{2}(\tau)$ for this
transition, which recovers~Eq.~(\ref{eq:Sat22Jul195618CEST2023}) for
the case~$P_\sigma=\gamma_1=\gamma_2$ that we considered previously.

There have been many studies of single-photon emission both
theoretically and experimentally, for which our approach is relevant
and that indeed prefigure our small~$N$ phenomenology.  This includes
other full quantum optical treatments (with a master equation) which
typically consider three-level systems~\cite{pegg86a,basche92a}, which
however do not report oscillations but merely a multi-exponential
return to uncorrelated emission.  There is indeed a range of
parameters that accommodate oscillations in~$g^{(2)}(\tau)$ and our
Erlang particular case fulfills them. Such oscillations are
conceptually noteworthy as they occur in a completely incoherent
system, exposing the time-ordering as the system liquefies. While the
parameter range is not particularly strict, it is easy to overlook if
one is not aware of such a possibility.  Elbows to antibunching have
also been observed in various systems, where they are often attributed
to intermediate states in a way that is compatible with our
mechanism~\cite{kitson98a,kurtsiefer00a,persson04a,neu12a,berthel15a,koperski18a,boll20a}. Descriptions
have been, however, based on rate equations and the consequences in
terms of better single-photon emission have been either ignored or
even considered problematic, the case of
Eq.~(\ref{eq:Sat22Jul164940CEST2023})
[Fig.~\ref{fig:Thu27Jul141838CEST2023}ii] being considered,
misguidedly, as an ideal.  Multilevel rate equations can provide
powerful results, including impressive accounts of~$g^{(2)}(\tau)$
over 11 orders of magnitude in time, for systems bathed in
highly-complex semiconductor environments featuring a plethora of dark
states and fluorescence intermittency~\cite{davanco14a}. Whether such
models compete or complement our mechanism, which endows photons with
qualitatively superior features of single-photon emission, remains to
be ascertained.  Our approach furthermore focused on the most basic
configuration to highlight the conceptual novelty, and should be
extended in several ways to tackle realistic experiments. Let alone
that the Erlang (degenerate) distribution is a very particular case
(with one parameter only whereas one naturally expect various
transitions to come with possibly largely varying parameters), the
number of steps in the cascade could also be a random variable, and
the possibility for various carriers to undergo such cascades
simultaneously be included. Our basic picture of liquefaction should
however remain and even apply to more general cases, e.g., that of
interactions between several emitters, that display curves similar to
Eq.~(\ref{eq:Tue25Jul142700CEST2023})~\cite{suarez19a}.

Our findings come with several conclusions. One is that the quest for
perfect single-photon sources has been so far driven by technological
improvements to reduce~$g^{(2)}(0)$ as much as possible from the basic
structure of a two-level system. This is a quantitative and asymptotic
race that is doomed to imperfection as the limitations are
fundamental: photodetection as a physical process will always detect
simultaneous photons from a two-level
system~\cite{lopezcarreno22a}. To obtain perfect single-photon
emission, one must open a gap somewhere~\cite{arXiv_khalid23a}. We
have provided a straightforward mechanism, furthermore of relevance in
solid-state platforms and that could otherwise be engineered, of a
cascade process that results in strong repulsions between photons with
the effect of imprinting strong correlations between them, similarly
to how interactions order and correlate molecules in a fluid. The
number~$N$ of cascades rules the magnitude of the effect and we have
given a general closed-form analytical expression for all~$N$. A
related conclusion is that~$g^{(2)}(0)$ itself is not the most
relevant measure for two-photon suppression. One must instead
consider~$g^{(2)}(\tau)$ locally around~$\tau=0$, and consider both
the power dependence of the short-time correlations as well as the
presence of oscillations or at least bunching (elbows) past the first
coherence time, as these mark the onset of short-time photon
ordering. This also suggest that considerably more types of quantum
lights are awaiting to be discovered and classified through such a
perspective. Finally, maybe the most far-reaching suggestion of our
approach, is that thermodynamic concepts that are central to describe
condensed matter could also provide more systematic and deeper
descriptions of quantum light, possibly relying on equations of states
to describe the various phases, as opposed to $n$-th order coherence
(almost always truncated to~$n=2$ anyway). Bunching, for instance,
might be related to plasma. Intriguingly, this condensed-matter
perspective would require to trade space for time.

\begin{acknowledgments}
  We thank S.~Kyu, F.~Sbresny, K.~Boos and K.~M\"uller for interesting
  discussions and feedback as this work was being conceived during a
  stay from the Authors in Munich. EdV acknowledges support from the
  CAM Pricit Plan (Ayudas de Excelencia del Profesorado
  Universitario), TUM-IAS Hans Fischer Fellowship and projects
  AEI/10.13039/501100011033 (2DEnLight) and Sin{\'e}rgico CAM 2020
  Y2020/TCS-6545 (NanoQuCo-CM).
\end{acknowledgments}

\bibliographystyle{naturemag}
\bibliography{sci,Books,arXiv}

\end{document}